\documentclass[aps,prb,twocolumn,showpacs,eqsecnum]{revtex4}

\usepackage{amssymb}
\usepackage{graphicx}
\usepackage{dcolumn}
\usepackage{bm}

\def\be{\begin{equation}}
\def\ee{\end{equation}}
\def\bea{\begin{eqnarray}}
\def\eea{\end{eqnarray}}

\begin{document}
\title{Non-linear thermoelectric transport: A class of nano-devices for high efficiency and large power output}
\author{Selman Hershfield, K. A. Muttalib and Bradley J. Nartowt}
\affiliation{Department of Physics, University of Florida, Gainesville FL 32611-8440}

\begin{abstract}

Molecular junctions and similar devices described by an energy dependent transmission coefficient can have a high linear response thermoelectric figure of merit. Since such devices are inherently non-linear, the full thermodynamic efficiency valid for any temperature and chemical potential difference across the leads is calculated.  The general features  in the energy dependence of the tranmission function that lead to high efficiency and also high power output are determined.  It is shown that
the device with the highest efficiency does not necessarily lead to large power output.  To illustrate this, we use a model called the t-stub model representing tunneling through an energy level connected to another energy level.  Within this model both high efficiency and high power output are achievable.  Futhermore, by connecting many nanodevices it is shown to be possible to scale up the power output without compromising  efficiency in an (exactly solvable) n-channel model even with tunneling between the devices.

\end{abstract}

\maketitle

\section{Introduction}

Thermoelectric materials \cite{book} can convert unused waste heat to electricity  (Seebeck effect) or use electricity for refrigeration (Peltier effect). A good thermoelectric material needs to have a good electrical conductivity $\sigma$, and at the same time a poor thermal conductivity $\kappa$. However, in normal bulk materials the two properties are related and follow the well-known Wiedemann-Franz law given by  $\kappa/\sigma T=\pi^2k_B^2/3e^2$, where $T$ is the temperature, $k_B$ is the Boltzmann constant, and $e$ is the electric charge. As a result it has not yet been possible to find bulk thermoelectric materials efficient enough to be cost effective except in specialized applications like space travel.

The subject has gained a lot of attention in recent years due to the increasing prospect of enhanced efficiency by nanostructural engineering \cite{review1,review2,majumdar,datta1,flensberg,buttiker,imry,sanchez,whitney,jordan}. It seems possible to control the electrical and thermal properties independently in such nanosystems \cite{dresselhaus,reddy}. The effectiveness of a thermoelectric material is usually estimated by its thermoelectric figure of merit $ZT \equiv S_e^2T\sigma/\kappa$, where $S_e$ is the thermopower or Seebeck coefficient,  and $\kappa$ contains contributions from electrons as well as phonons. Currently, best materials have $ZT \sim 1$, while it is estimated that $ZT > 3$ would be industrially competitive \cite{majumdar}. 

Mahan and Sofo \cite{mahan} considered the optimization of the figure of merit as a mathematical problem and found that for an ideal  delta-function distribution of the transmission function $\mathcal{T}(E)$ as a function of energy $E$, the figure of merit diverges in the absence of any phonon contribution to the thermal conductivity \cite{majumdar,hochbaum,boukai}. It was argued later \cite{murphy} that a molecular junction can also give rise to a diverging figure of merit. Other theoretical models have also predicted large $ZT$ values for nanosystems, e.g. for double quantum dots \cite{ws}. While there is reason to be optimistic about the prospects for making
useful nanostructure thermoelectric devices, there are a number of 
major issues which need to be addressed.  In this paper we address
three of them.

1. The figure of merit for bulk systems, $ZT$, is derived in the linear
response regime.  This is quite reasonable in bulk systems because one
is expanding in the gradient in the temperature and the electrical 
potential.  It is possible to have large temperature differences across
a sample and yet small gradients in temperature within the sample.
For the type of nanostructure considered in the above and in this paper
the temperature and electrical potential gradients occur on the 
nanometer scale, leading to enormous gradients.  Thus, in nanostructures
the interesting regime for extracting energy is the nonlinear response
regime.  What is the response of high $ZT$ nanostructure devices
in the nonlinear response regime?

2. Also, from bulk systems one would expect that a higher 
thermodynamic efficiency would lead to a higher power output.
We will show in the next section that there are some models
for nanostructures which lead to the maximum thermodynamic
efficiency in the limit where the power output goes to zero.  
This is possible because the efficiency, $\eta$, is the ratio
of the electrical energy extracted to the heat
transfer between reservoirs.  This ratio can approach the Carnot limit
maximum as both terms in the ratio go to zero.  Is it possible to 
achieve large power output and high efficiency simultaneously in
nanostructured devices?

3. In nanostructures the currents and heat transfers are very small
compared to macroscopic electrical currents and heat transfers.
To make nanostructure devices useful for extracting energy on the 
macroscopic scale, one needs to scale up the response.  The obvious
way to do this is to put many nanoscale devices in parallel.
If these devices are very far apart, then clearly the power output
scales with the number of devices.  However, as the devices are
put more closely together to optimize the power output per unit area,
there will eventually be tunneling between devices.  How does tunneling
between nanoscale thermoelectric devices effect their efficiency
and power output?

While there have been electronic structure and transport calculations
through molecular junctions\cite{datta1,finch,liu}, in this paper
we consider a model called the t-stub \cite{stadler}, which has the important
features of more specific calculations such as a rapidly varying
transmission coefficient near the Fermi energy \cite{abbout}.  
This model has been used to understand
tight binding and density functional theory calculations of 
tunneling through molecules and has even been parametrized for 
specific molecules.  
The t-stub model is closely related to models used for interference
in wave guides, and interference is the mechanism that produces
the rapidly varying transmission coefficient.   While more realistic models of 
granular semiconductors \cite{tripathi,glatz} have been studied in the linear response regime,
the t-stub model is suited toward answering the
general questions above about nonlinear response, optimizing
power and efficiency, and tunneling between nanoscale devices.
Within this model we will still be able to estimate and compare the
power output of nanoscale devices to present commerical devices.

The rest of the paper is organized as follows.  In Sec. II we present
the formalism for doing non-linear response and obtain from thermodynamic
arguments the criteria for obtaining both a large power output and
a high efficiency in the non-linear response regime.
In Sec. III the t-stub model is solved in the nonlinear transport
regime. It is shown that for the parameters chosen based on the insight
developed from Sec. II one can obtain large efficiency and power output simultaneously.
In Sec. IV the effect of coupling many t-stub devices in parallel is
caclulated as a function of the number of devices coupled.  
In Sec. V we discuss the inclusion of phonon contributions to the
thermal conductivity, and in Sec. IV we summarize our findings for the
three questions posed in this introduction.  Some technical details 
are presented in the Appendices.

\section{Efficiency and Power Output for Non-linear Response}

\begin{figure}
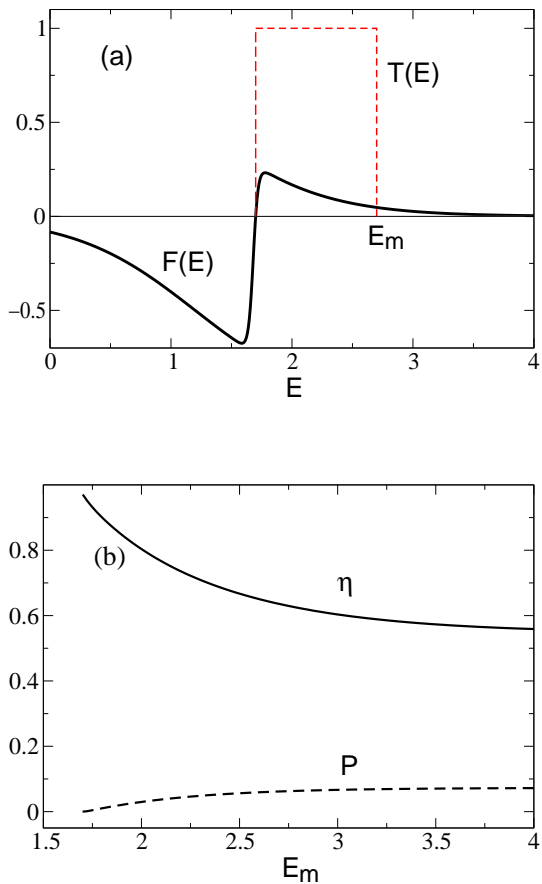


\includegraphics[angle=270, width=0.3\textheight]{Fig-1a.eps}
\newline
\newline
\newline

\includegraphics[angle=270, width=0.3\textheight]{Fig-1b.eps}
\newline
\newline

\caption{\label{fig-FofE}(Color online)  Optimizing transmission function $\mathcal{T}(E)$ in the non-linear regime: (a) Difference of the Fermi functions $F(E)\equiv f_L(E)-f_R(E)$ (solid black curve)  with  $T_L=0.5$, $\mu_L=1.0$, $\mu_R=1.68$, and $\hat{E}=1.7$ and a square-wave   $\mathcal{T}(E)$  (dashed red curve) starting at $E=\hat{E}$ and ending at some arbitrary $E_m$. Note that here $\hat{E} > \mu_R > \mu_L$. (b) The efficiency $\eta$ (solid line) and the power output $P$ (dashed line) as functions of the parameter $E_m$. Note that the power output is zero where the efficiency is maximum.}
\end{figure}

A thermodynamic heat engine takes heat $Q_L$ from a reservoir kept at temperature $T_L$, does work $W$, and releases heat $Q_R$ to a  reservoir kept at a lower temperature $T_R$. The efficiency is defined as the ratio of work done to the heat extracted from the high temperature reservoir:
$\eta\equiv W/Q_L=1-Q_R/Q_L$,
where the latter follows from the conservation of energy. The power output, on the other hand, is the product of the charge current times the voltage drop across the device.    In our notation it is given by $P = (\mu_R-\mu_L)I_N$ where $I_N$ is the number current and $\mu_L$ and $\mu_R$ are the chemical potentials of the left and the right leads, respectively.  In the following we fix $\mu_L$,  and $\mu_R$ is determined by the load connected to the thermoelectric generator. In terms of an energy dependent transmission function $\mathcal{T}(E)$ the power output can be written as 
\bea
&& P =  \frac{1}{h} (\mu_R-\mu_L)\int dE \mathcal{T}(E)F(E);\cr
&& F(E)\equiv  f_L(\mu_L,T_L; E)-f_R(\mu_R,T_R; E)
\label{eq-power}
\eea
where $f_j(\mu_j,T_j;E)\equiv  1/(1+e^{(E-\mu_j)/k_BT_j})$ are the Fermi functions in the two leads.
Using this notation
the efficiency can be written in terms of the transmission function as
\bea
\eta=\frac{(\mu_R-\mu_L)\int dE \mathcal{T}(E)F(E)}{\int dE (E-\mu_L) \mathcal{T}(E)F(E)}.
\label{eq-eta}
\eea
This expression, together with Eq.~(\ref{eq-power}), allows us to optimize the efficiency as well as the power output by carefully matching $\mathcal{T}(E)$ for a given $F(E)$. 
Figure \ref{fig-FofE} shows an example of $F(E)$ for an arbitrarily chosen set of parameter values (in units of $\mu_L=1$, $h=1$) as shown in the figure caption.

Note that $F(E)$ crosses over from negative to positive values at $E=\hat{E}$ obtained from
$
(\hat{E}-\mu_L)/T_L= (\hat{E}-\mu_R)/T_R. 
$
Solving for $T_R$ gives
\be
T_R=T_L\frac{\mu_R-\hat{E}}{\mu_L-\hat{E}}; \;\;\; F(\hat{E})=0.
\label{eq-TR}
\ee
Thus one sees that for a given $T_L$ and $T_R$ with $0 < T_R < T_L$, both chemical potentials must lie on the same side of the parameter $\hat{E}$, satisfying the inequalities 
$\hat{E} \ge \mu_R \ge \mu_L$ (case I) or $\mu_L \ge \mu_R \ge \hat{E}$ (case II).
For definiteness, and without loss of generality, we will only consider case I in all our examples and discussions.

It might seem from Eqs.~(\ref{eq-power}), (\ref{eq-eta}) that one can simply increase the power output $P$, and hence the efficiency $\eta$, by arbitrarily increasing the difference in chemical potentials between the two leads. However, the number current $I_N$ is maximum when the generator is in `short circuit' (without any external load resistance) and $\mu_R=\mu_L$, while $I_N\to 0$ when $\mu_R-\mu_L = \Delta\mu_0$ where $\Delta\mu_0$ depends on the energy dependence of the transmission function. (Beyond $\Delta\mu_0$, the current changes sign and the device takes in energy rather than generating it.) The power output $P$ (and hence the efficiency $\eta$) is therefore zero at both these limits. The choice of $\mu_R$ that maximizes $P$ within these limits is usually quite different from the choice that maximizes $\eta$. As a simple soluble example, let us consider $\mathcal{T}(E)\propto \mathcal{T}_0\delta(E-E')$, $E' > \hat{E}$, leading to $P\propto (\mu_R-\mu_L)\mathcal{T}_0F(E')$ and $\eta= (\mu_R-\mu_L)/(E'-\mu_L)$. Then $P$ is maximized if $E'$ is chosen to coincide with the maximum of $F(E)$, while $\eta$ is maximized if $E'$ is chosen to be equal to its smallest allowed value, $\hat{E}$. Note that if $E'=\hat{E}=\mu_R$ exactly, one gets the ideal efficiency $\eta=1$. This is consistent with the result that an ideal delta function form of the transmission function can lead to a divergent figure of merit (in the absence of phonons) \cite{mahan}. However for this choice, one has $F(E=\hat{E})=0$, which implies $P=0$. This is an ideally efficient but completely useless generator. This is why considering the efficiency (or the figure of merit) without considering the power output can be misleading.  

The general form of $F(E)$ shown in Figure \ref{fig-FofE} provides significant insights into the features of $\mathcal{T}(E)$ that could be helpful to optimize both efficiency and power. For example,  for a given $F(E)$, one can minimize the \textit{cancellations} from positive and negative parts of $F(E)$ contributing to $P$ if $\mathcal{T}(E)$ is chosen to have negligible weight in the entire range where $F(E) <0$. (Note that $\mathcal{T}(E)$ can not be negative, and we only consider case I. For case II, a similar condition would mean having $\mathcal{T}(E)\to 0$  in the entire range where $F(E) >0$.) 
This insight, together with the results provided by the delta-function model considered above, immediately suggest certain design criteria for a good thermoelectric material. First, it should have a tunable phenomena leading to a negligible value for $\mathcal{T}(E)$ in a range of $E$ dictated by $F(E)$. For example, a square $\mathcal{T}(E)$ starting at $E=\hat{E}$ and ending at some arbitrary $E_m$ as shown in Figure \ref{fig-FofE} avoids the negative parts of $F(E)$ and at the same time takes advantage of the maximum of $F(E)$.
Second, any design has to optimize the power output and the efficiency simultaneously, as opposed to maximizing one or the other. In the example of Figure \ref{fig-FofE}, the power output $P$ is zero where $\eta$ is maximum, and as a function of the width of the square transmission function the power increases  while the efficiency decreases \cite{whitney1}. Third, we note that while the `strength parameter' $\mathcal{T}_0$ in the delta-function model drops out of the efficiency, it directly increases the power output. Thus it should be possible to optimize $\eta$ and $P$ for a single channel device, while the power output can subsequently be made large  by `scaling up' the number current by increasing the number of channels. In the following we will first consider a single chain model that allows us to tune the energy dependence of the transmission function in a desirable way, and then discuss possible ways of scaling up the power output.

While in this work we will consider the full non-linear regime, the connection of the thermodynamic efficiency $\eta$ with the figure of merit $ZT$ mentioned in the Introduction (and valid only in the linear response regime) is briefly discussed in Appendix I. In particular, the goal of $ZT\ge 3$  can be rewritten as $\eta \ge 0.3 \times \Delta T/T$ where $T$ is the average temperature, and linear response regime implies $\Delta T/T \ll 1$. In order to be able to make a valid comparison, we will define  $\eta = \eta_c \times \bar{\eta}$   where $\eta_c$ is the Carnot efficiency. Thus an industrially competitive $\eta$ would mean 
\be 
\eta/\eta_c \ge 0.3; \;\;\; \eta_c\equiv 1-\frac{T_R}{T_L}.
\label{goal}
\ee
In the non-linear regime $\eta_c$ need not be much smaller than unity. In the context of space travel the temperature differences can be quite large with $T_L \gg T_R$ and $\eta _c \approx 1$. 
However for harnessing waste energy on the earth, we will need $T_R$ to be the room temperature  and $T_L \sim 450$ K, approximately the temperature of a running automobile engine. In other words, the goal is not only to have  $\eta/\eta_c \ge 0.3$, but also to have it for $\eta_c\sim 1/3$. We will see below that our model achieves both.

\section{A Model System}

Although a square-wave $\mathcal{T}(E)$ considered in Figure 1 would be ideal, it is not clear how such a shape can be obtained in practice in a nano-system given the fact that any tunnel-barrier designed to cut-off the transmission in a desired energy range would have its own inherent interference effects that would destroy the sharpness of the cutoff. Our goal here is to take advantage of the interference effcts in producing a $\mathcal{T}(E)$ as close to the square-wave as possible, keeping in mind that the device has to be geometrically scalable to increase the power output.
For these reasons, it is more convenient to start with a simple exactly solvable  microscopic model which has the potential to achieve any desired $\mathcal{T}(E)$ and which, at the same time, is geometrically scalable.

\begin{figure}
\includegraphics[angle=0, width=0.3\textheight]{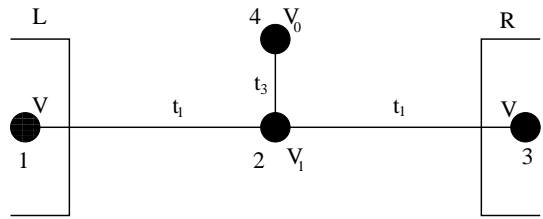}
\caption{ One dimensional model consisting of a chain connected to a central site $2$ and an extra site $4$ with site energies $V_1$ and $V_0$, respectively. The hopping parameter between sites 2 and 4 is $t_3$. Site $2$ is connected to site $1$ in the left lead $L$ and to site $3$ in the right lead $R$ with hopping parameters $t_1$. Leads $L$ and $R$ are characterized by hopping parameters $t$ and site energies $V$. The site $4$ is not directly connected to any of the leads. We will refer to this as the t-stub model.}
\label{fig-toy-phonon}
\end{figure}

A single chain model with two channels, one purely electronic and another phonon-assisted, is expected to show a dip in the electron transmission as a function of energy due to destructive interference between the channels. This idea leads to the simplest model that allows us to tune the energy dependence of $\mathcal{T}(E)$.
Consider the model shown in Figure \ref{fig-toy-phonon}
where an isolated chain is attached to an extra site on the side (site 4 in
Fig. 2). 
This is known in the literature as the $t$-stub model, and as noted in the Introduction, has been used to make connections with realistic molecular junctions \cite{stadler}. 
The extra site may be regarded as either an energy level within the same
atom as site 2 in Fig. 2, or an energy level on a neighboring atom.
In the case when the occupancy of site 2 is small, it may also be regarded
as corresponding to a single virtual phonon excitation.  This later analogy
breaks down when the occupancy of site 2 is not small, which is the case
considered here. 
The Hamiltonian $H$ of the isolated chain is given by 
\bea
H=\left( \begin{array}{cccc} V &  t_1 & 0 & 0 \\
t_1 & V_1 & t_1 & t_3 \\ 0 & t_1 & V & 0 \\ 0 & t_3 & 0 & V_0 \end{array} \right). 
\eea
The retarded Green function with the leads is   
\bea
G^R = \left[EI-H-\Sigma\right]^{-1}
\eea
where 
$\Sigma_{11}=\Sigma_L$, $\Sigma_{33}=\Sigma_R$
and all other $\Sigma_{ij}=0$. The `self energies' $\Sigma_{L,R}$ are due to coupling to the left and the right leads, respectively \cite{datta},
and are given by 
\be
\label{Sigma-el}
\Sigma_{L,R}=t^2g_{L,R}=-te^{ika} 
\ee
where $t$ is the hopping element in the leads. Here $g_{L,R}$ is the surface Green function of the left or
right semi-infinite lead, $k$ is the incident wave vector and $a$ is the lattice constant in the two leads and we have assumed symmetric leads. The tight binding model in the leads correspond to
$V=2t$ and $E=2t(1-\cos ka)$, so that the bandwidth is $\mathcal{W}=4t$. In the following in all our examples, we will always choose our energy parameters in units of $t$.

The Green function across the chain $G_{13}$ is 
\bea
G_{13} = \frac{t^2_1}{t^2}\frac{\Sigma_L\Sigma_R}{t^2(E-V_1-\Sigma_2)-t_1^2(\Sigma_L+\Sigma_R)}
\eea
where $\Sigma_2 \equiv  t_3^2/(E-V_0)$.
For symmetric leads, defining 
$
a_0\equiv E-V-\Sigma_{L,R}=\frac{|t|^2}{\Sigma_{L,R}},
$
we can rewrite it as 
\be
G_{13}=\frac{t^2_1E_0}{a_0D_1}\equiv g_0,
\label{G13}
\ee
where
$D_1\equiv a_0(E_0E_1-t_3^2)-2E_0t_1^2$, $E_0\equiv E-V_0$ and $E_1\equiv E-V_1$.
The transmission coefficient $\mathcal{T}$ is then given by 
\bea
\label{Tdef}
\mathcal{T} &=& v_Lv_R|G^R_{13}|^2 = E(4t-E)|g_0|^2 ,
\eea
where $v_L$, $v_R$ are the left and right channel velocities, respectively, given by $v_L=v_R=2t\sin ka=\sqrt{E(4t-E)}$. 
Since $g_0\propto E_0$, this exhibits a zero at the resonant energy $E=V_0$, in addition to the zeros at the band edges.   Note that the parameter $V_0$ allows us to tune the position of the dip, while $V_1$ and $t_3$ can be used to vary the width. To some extent, these molecular parameters can be tuned by an electric field \cite{park} or possibly by an external gate voltage. In particular, the parameters can be chosen to generate a $\mathcal{T}(E)$ which has the desired feature of being negligible for a range of $E$ where $F(E)$ is negative. 
 In order to avoid any artificial  effects from the band edges, we will always use thermodynamic parameters such that $F(E)$ is negligible at the upper band edge. Since $\mathcal{T}(E)$ for our choice of $t_3$ is negligible at the lower band edge, the product $\mathcal{T}(E)F(E)$, and hence the resulting power and efficiency, will be largely insensitive to the cut-off in the model at either band-edge.

\begin{figure}[!htp]
\includegraphics[angle=270, width=0.3\textheight]{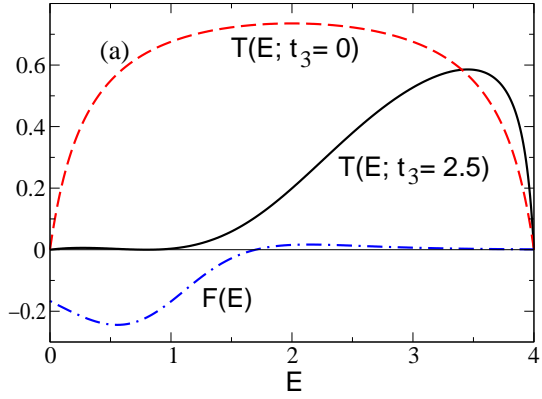}
\newline
\newline
\newline

\includegraphics[angle=270, width=0.3\textheight]{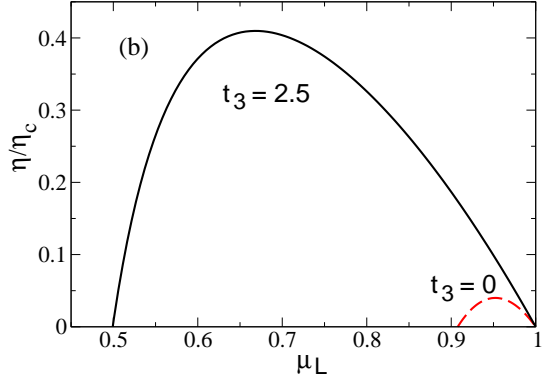}
\newline
\newline
\newline

\includegraphics[angle=270, width=0.3\textheight]{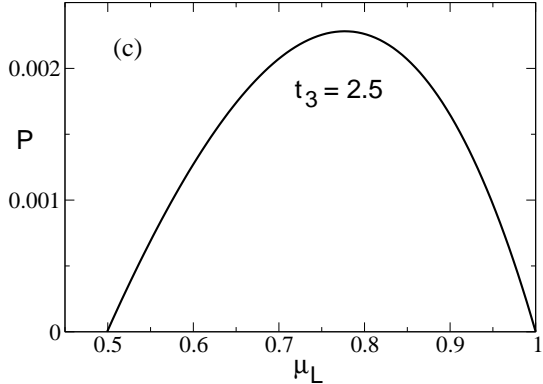}
\newline
\caption{(Color online) The single chain t-stub model: (a) Transmission function $\mathcal{T}(E)$ with the extra side level (solid black curve), corresponding to $t_3=2.5$ and without the extra side level (dashed red curve), corresponding to $t_3=0$, all in units of $t$ (bandwidth $\mathcal{W}=4$). The parameters used for both cases are $t_1=1$, $V_0=V_1=0.8$. For comparison, the blue dash-dotted  curve shows $F(E)$ for $\mu_L=0.65$, $T_L=0.5$, $\mu_R=1.0$ and $\hat{E}=1.7$ such that $\mu_L < \mu_R < \hat{E}$. (b) The efficiency $\eta/\eta_c$ corresponding to the $F(E)$ and the two $\mathcal{T}(E)$ shown in (a), as functions of $\mu_L$: red dotted line corresponds to the case $t_3=0$ and the black solid line is for $t_3=2.5$. Note the order of magnitude increase in efficiency showing the importance of the interference effects from the extra side level. (c) Power output $P$ (in units of $t^2/h)$ corresponding to the $\mathcal{T}(E)$ in (a) with $t_3=2.5$.} 
\label{fig-TofE-SingleChain}
\end{figure}

Now we show the importance of the extra site. In the example shown
in Figure \ref{fig-TofE-SingleChain} we  compare two cases, one with  $t_3=0$ (no side level) and the other with $t_3=2.5$ (in units of $t=1$), together with a given choice of $F(E)$. 
We have chosen $V_0$ and $V_1$ to produce a negligible $\mathcal{T}(E)$ for $E<\hat{E}=1.7$ for the choice $t_3=2.5$.
The resulting efficiencies $\eta/\eta_c$, for fixed values of $T_L=0.5$ and $\mu_R=1.0$ are shown in the middle panels (a) and (b) of Figure \ref{fig-TofE-SingleChain} as a function of $\mu_L$, where we keep $\hat{E}=1.7$ fixed in order to take advantage of the feature in $\mathcal{T}(E)$ in the top panel. The range of $\mu_L$ is restricted, in each case, by the requirement that $I_N > 0$.  Note that while the maximum $\eta/\eta_c$ without the side level is only $\eta/\eta_c\approx 0.04$ (red dashed line), it can be one order of magnitude larger for the finite value of $t_3$ chosen here (black solid line). 
Clearly, the matching of $\mathcal{T}(E)$ with $F(E)$, tuned with the help of the interference associated with the side level, can be an effective tool to increase the efficiency of a nano-engineered thermoelectric material.
Indeed when compared with Eq.~(\ref{goal}), the increased efficiency in the above example exceeds the threshold for industrial competitiveness. Moreover the maximum value $\eta/\eta_c\approx 0.4$ occurs for $\mu_L=0.65$, for which $T_R/T_L=2/3$ or $\eta_c=1/3$. As mentioned earlier, this fulfills the requirement for a practical device to harness waste heat energy from the environment.

As for the power $P$ it is important to note that although, as warned before, the maximum of $P$ and $\eta$ do not occur for the same value of $\mu_L$, there is a range of $\mu_L$ for which $\eta/\eta_c > 0.3$ and $P > 0.001$ (in units of $t^2/h$) simultaneously. However,  $P\sim 10^{-3}$ is unacceptable as a practical device.  
For example $T_R=300 K$ in the above example of Figure \ref{fig-TofE-SingleChain} corresponds to $t = 3k_BT_R$. This implies $P=10^{-3}\times t^2/h \sim 10^{-10}$ Watts. 
Comparing with currently available bulk commercial devices \cite{selman} with $P\approx 4000$ Watts/m$^2$ for the power per unit area of the thermoelements,  it is clear that it is absolutely essential to be able to geometrically `scale up' the model in order to obtain the necessary power output.  

\begin{figure}
\includegraphics[angle=270, width=0.3\textheight]{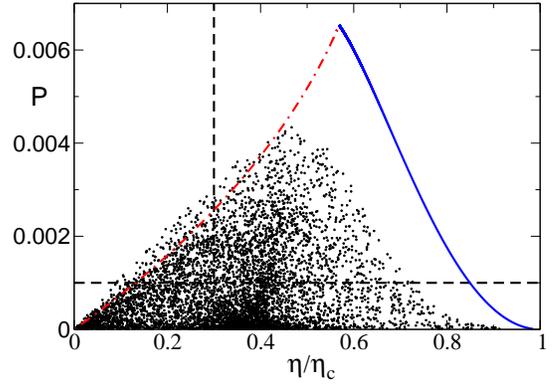}
\caption{Efficiency and power for randomly chosen $3\times 10^3$ sets of parameters within the range $t_1=1\pm 1, t_3=2.5 \pm 1$ and  $V_0,=V_1=0.8 \pm 2$, keeping the thermodynamic parameters $\mu_R=1.0$ and $\hat{E}=1.7$ fixed and choosing $\mu_L=0.65$. The dashed lines demarcate the parameters for which one can have $\eta/\eta_c > 0.3$ and $P > 0.001$ simultaneously.  For comparison, the blue solid line shows the maximum $P$ for a given $\eta/\eta_c$ as obtained for an ideal square-wave between $\hat{E}$ and $E_m$ as shown in Figure 1, for $\hat{E} <E_m<4t$. The red dash-dotted line corresponds to the case where the square-wave is extended into the negative part of $F(E)$. Note that in this case, the square-wave is no longer the limiting envelop.  }
\label{montecarlo}
\end{figure}

We emphasize that the values of the microscopic model parameters chosen for illustration in Fig. \ref{fig-TofE-SingleChain} are not the `best' (fine-tuned) values. In fact, while we have not explored the entire parameter space systematically, a simple parameter sampling of $3\times 10^3$ possible sets of the parameters around the chosen values, as explained in Figure \ref{montecarlo}, shows two important features. 
First, our results are quite generic in the sense that typically a variety of different combinations of the parameters will give similar values of $\eta$ and $P$. Second, fine-tuning the parameters could actually increase $\eta$ significantly 
towards the maximum envelop obtained for an ideal square-wave form of $\mathcal{T}(E)$, shown by the blue solid line. In the square-wave case, starting from the Mahan-Sofo limit of ideal efficiency and zero power for $E_m\to \hat{E}$, $P$ increases (and $\eta/\eta_c$ decreases) as $E_m$ increases, the maximum of $P= 0.0065$ occuring for $E_m=4t$ (the band edge), corresponding to the maximum width for which $F(E)$ is positive. Increasing the width of the square wave any further requires including the negaive part of $F(E)$ which, as discussed in Sec II, decreases $P$ significantly in the region covered by the red dash-dotted line in Figure \ref{montecarlo}. Clearly the square-wave no longer corresponds to an optimum envelop in this regime since reducing the ideal transmission of a square-wave in the negative $F(E)$ regime with any other shape would lead to an increase in $P$.

While the density of the points in Figure \ref{montecarlo} does not have any meaning, it is clear that a wide range of parameters is available where $\eta/\eta_c > 0.3$ and $P > 0.001$ simultaneously.  Nevertheless, the maximum of $P$,  even for an ideal square-wave, remains unacceptably small, $P \approx 0.007$, unless it can be geometrically scaled up.
In the following section we extend the model to $n$ number of coupled chains, which turns out to be still exactly soluble.

\section{The scaled up model: $n$ chains}

\begin{figure}[!htp]
\includegraphics[angle=0, width=0.3\textheight]{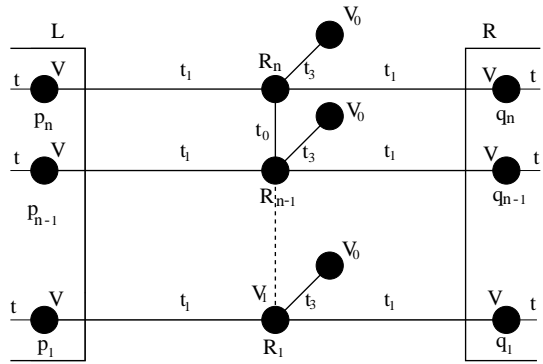}
\caption{The scaled up quasi-1d chain, connecting site $R_j$ with energy $V_1$ of one chain to the corresponding site $R_{j\pm 1}$ in the neighboring chain, with hopping parameter $t_0$, leaving the bond $t_3$ as the only connection to the $V_o$ site. The parameters for each individual chain as well as the left and right leads $L$ and $R$ are the same as in Figure \ref{fig-toy-phonon}.}
\label{fig-toy-phonon-scaled}
\end{figure}

We extend the single chain $t$-stub model to a system with $n$ number of chains, each consisting of a `left' and  a `right' site $p_k$ and $q_k$, respectively, with site energies $V=2t$, and a `middle' site $R_k$ with site energy $V_1$. The lead-junction hopping parameters are given by $V_{p_kR_k}=V_{R_kq_k}=t_1$.  Each site $R_k$  has a side level with hopping element $t_3$ connected to a site with energy $V_0$, as shown in Figure \ref{fig-toy-phonon-scaled}.
Clearly, if the chains are independent, the total power would simply scale with the number of chains. However, in a nano-system, two chains nearby are always coupled due to quantum tunneling via nearby atoms, and it is not clear if the interference effects in the single chain would survive in the presence of multiple possible paths generated by interchain couplings. Here we will consider the simplest case where
the chains are connected at sites $R_k$ with hopping parameters $V_{R_kR_{k+ 1}}=V_{R_kR_{k-1}}=t_0$.
We will start with $n-1$ chains and call the Green functions $G^0_{p_jq_k}$ for arbitrary $j, k\le n-1$. We will then add the $n$th chain (with site $R_n$ connected to site $R_{n-1}$ by hopping parameter $t_0$) and evaluate the resulting $G_{p_jq_k}$ recursively. From symmetry, we will only need $G_{p_jq_k}$ for $j\le k\le n$.

In order to evaluate $G_{p_jq_k}$ for $j\le k\le n$ we will need the `building blocks' $G_{R_jR_k}$, $G_{p_jR_k}$ and $G_{R_jq_k}$. To start with, we note that $G_{R_nR_n}$ satisfies the recursion relation 
\bea
\label{recXn}
X_n=\frac{1}{1-bX_{n-1}}; \;\;\; X_n\equiv \frac{D_1}{a_0E_0} G_{R_nR_n},
\eea
where we defined 
\be
\label{b}
b\equiv \left(\frac{a_0E_0t_0}{D_1}\right)^2.
\ee
Note that $X_1=1$. The solution to Eq. (\ref{recXn}) is given by \cite{mathematica}
\bea
\label{solnXn}
X_n=\frac{2[(1+\alpha)^n-(1-\alpha)^n]}{(1+\alpha)^{n+1}-(1-\alpha)^{n+1}}; \;\;\;\alpha\equiv \sqrt{1-4b}.
\eea
Now consider $G_{R_nq_k}$, which satisfies a recursion relation
\be
\label{recYnk}
Y_{n,k}=\sqrt{b}X_nY_{n-1,k}; \;\;\;Y_{n,j}\equiv G_{p_jR_n}=G_{R_nq_j}; \;\;\; j < n.
\ee
The solution to Eq.(\ref{recYnk}) is given by
\be
Y_{n,k}=\frac{E_0t_1}{D_1}X_k\prod_{m=k}^{n-1}\sqrt{b}X_{m+1}; \;\;\; k<n.
\ee

We are now in a position to evaluate the Green functions across the chain. In terms of the $X$ and $Y$ functions, the recursion relations eventually lead to the following expressions:
\bea
G_{p_kq_k}
&=&  \frac{E_0t^2_1}{a_0D_1}[1+bX_kX_{k-1}]+ t_0\sum_{m=k}^{n-1}Y_{m+1,k}Y_{m,k};\cr
&\;& k < n-1
\label{GpkqkXY}
\eea
and 
\bea
G_{p_jq_k}
&=& \frac{t_0E_0t_1}{D_1}X_kY_{k-1,j}+ t_0\sum_{m=k}^{n-1}Y_{m+1,k}Y_{m,j}; \cr
&\;& j < k < n-1.
\label{GpjqkXY}
\eea
Note that the transmission involves $\sum |G_{p_jq_k}|^2$, so the above expression as a sum over sites gets very complicated. However
it turns out that by using the recursion relation (\ref{recXn}), it is possible to rewrite them as products instead of sums. In particular, one gets
\bea
\label{GpkqkX}
G_{p_{k}q_{k}}=g_0 \frac{X_{n}X_{n-1}\cdots X_{k}}{X_2X_3\cdots X_{n-k}}.
\eea 
The proof is given in Appendix II.
Given the above,
the expression for $G_{p_jq_k}$ can be simplified as
\bea
\label{GpjqkX}
G_{p_jq_k}  
=Q_{j,k}G^n_{p_kq_k}; \;\;\;Q_{j,k} \equiv  (\sqrt{b})^{k-j}\prod_{m=j}^{k-1} X_m.
\eea

Defining  
$z \equiv (1-\alpha)/(1+\alpha)$,
the expressions for the Green functions can be rewritten as 
\bea
G_{p_kq_k} &=& \frac{g_0}{\alpha}\frac{(1-z^k)(1-z^{n-k+1})}{(1-z^{n+1})}; \;\;\;\cr
G_{p_jq_k} &=&  \frac{g_0}{\alpha}\frac{z^{(k-j)/2}}{1-z^{n+1}}(1-z^j)(1-z^{n-k+1}) 
\label{Gz}
\eea
so that the transmission function becomes:
\bea
\label{TEalln}
&\;&\mathcal{T}(E) = v^2\left[\sum_{k=1}^{n}|G^n_{p_kq_k}|^2 + 2 \sum_{k=2}^{n}\sum_{j=1}^{k-1} |G^n_{p_jq_k}|^2\right]\cr
&=& \mathcal{T}_n[\sum_{k=1}^{n}|1-z^k|^2|1-z^{n-k+1}|^2 \cr
&+&  2\sum_{k=2}^{n} |1-z^{n-k+1}|^2 \sum_{j=1}^{k-1}|z|^{k-j}|1-z^j|^2]
\eea 
where $v$ is the channel velocity
and 
\bea 
\label{Tn}
\mathcal{T}_n \equiv v^2\frac{|g_0|^2}{|\alpha|^2|1-z^{n+1}|^2}.
\eea 
This is the exact result for the transmission function of the $n$-chain model.
It is possible to sum the terms analytically, but for a given finite $n$ it is easier to evaluate them directly numerically.

\begin{figure}
\includegraphics[angle=270, width=0.3\textheight]{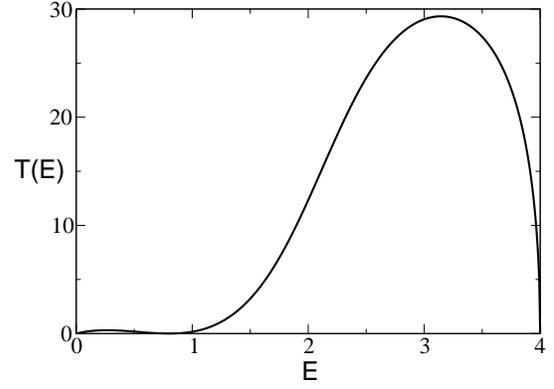}
\newline
\caption{Transmission $\mathcal{T}(E)$ 
for identical set of parameters as in the single chain model shown in Figure \ref{fig-TofE-SingleChain}, with interchain coupling $t_0=1$ and the number of chains $n=50$. Maximum efficiency remains similar to the single chain model while the maximum power output scales with $n$.}
\label{fig-TofE-n}
\end{figure}

Figure \ref{fig-TofE-n} shows  evaluation of $\mathcal{T}(E)$ from Eq. (\ref{TEalln}) for identical single chain parameters as used in Figure \ref{fig-TofE-SingleChain} (namely  $t_1=1$, $t_3=2.5$ and $V_1=V_0=0.8$), and using the interchain hopping parameter $t_0=1$, with $n=50$. 
Note that the maximum of $\mathcal{T}(E)$ for $n=50$ is almost a factor $50$ larger compared to $\mathcal{T}(E)$ for $n=1$ shown in Fig \ref{fig-TofE-SingleChain}, and it has the same features helpful for obtaining a large efficiency. Thus essentially $\mathcal{T}(E)$ scales with $n$, at least for the chosen values of the parameters. 
For comparison with the single chain case we choose the same thermodynamic parameters $\mu_R=1.0$, $\hat{E}=1.7$ and $\mu_L=0.65$, which gives $\eta/\eta_c=0.43$ and $P=0.1$.  
Thus while the efficiency remains undiminished (in fact it is slightly enhanced), the power output scales with $n$ as expected. In other words, it is possible to scale up the single chain power without compromising the efficiency, by simply increasing the number of chains. 

The estimate that $\mathcal{T}(E)$ and hence $P$ essentially scales with $n$ for large $n$ can be understood from 
a simple continuum ($n\to \infty$) limit. As shown in Appendix III, the transmission in the continuum limit is given by
\bea
\mathcal{T'}(E)=v^2\left|\frac{g_0}{\alpha}\right|^2
\label{Tinfinity}
\eea
where $v$ is the channel velocity defined earlier. This is the same transmission function per channel obtained in Eq.~(\ref{TEalln}), in the limit $z\to 0$. In this limit all channels become independent (renormalized by the parameter $\alpha$), and the total transmission simply scales with the number of channels $n$.

Going back to the single chain estimate for power and using the fact that scaling with $n$ holds for large $n$, we see that by putting chains $10$ nm apart and connecting them by the cross-bond $t_0$, it should be possible to achieve a power output of $P\sim 10^{-10}$ Watts/(10$^{-8}$ m)$^2$ $\sim 10^{6}$ Watts/m$^2$ (keeping the efficiency near $\eta/\eta_c > 0.4$). This is several orders of magnitude larger than the bulk thermoelectric generators currently available commercially when measured as power per unit area of the thermoelements.
Commercial devices can have their power output increased further by using essentially non-planar interfaces to increase the effective area between hot and cold regions \cite{selman}.  This could also be done for the devices based on the scaled up model.

We emphasize here again that in addition to choosing the values of the microscopic model parameters $t_1, t_3, V_0, V_1$ as those of the single channel case, the connecting bond hopping parameter $t_0=1$ has been chosen as the simplest possibility  and not as the `best' fine-tuned value. 
As in the single channel case, we expect that it should be possible to increase the efficiency further by fine-tuning the parameters.
The important point is that there would be   
a wide range of microscopic as well as thermodynamic parameters that can yield $\eta/\eta_c > 0.3$, and power $P> 0.001 \times n$, simultaneously. While we have chosen $\eta_c=1/3$ corresponding to a high temperature bath with $T_L=450$ K, it is clear that the design of a practical industrially competitive thermoelectric device should be possible even with a lower $T_L$.

\section{Phonons in a single chain}

In practice, the denominator of Eq.~(\ref{eq-eta}) should have an added contribution from phonons that would decrease the efficiency. In analogy to the electronic contribution $I_e = \int dE (E-\mu_L) \mathcal{T}(E)F(E)$, the phonon contribution to the energy flux can be written as
\bea
I_{ph} &=& \frac{1}{2\pi}\int_0^{\infty} dE\;E \xi(E) B(E); \cr 
B(E)&\equiv & \eta_L(E)-\eta_R(E)
\eea
where $\eta(E)$ is the Bose distribution function
$
\eta (\omega)= 1/[\exp(\hbar\omega/k_B T)-1],
$
and the subscripts $L$ and $R$ refer to the left and right leads, respectively. Here $\xi(E)$ is the phonon transmission function. For phonons in the leads with dispersion relation $ \omega^2=\omega^2_0(1-\cos Ka)$ where $K$ is the phonon wave vector, the transmission function can be expressed as 
\bea
\xi(\omega) & \equiv & \rm{Tr}[\Lambda_R(\omega)\mathcal{G}_{1,3}(\omega)\Lambda_L(\omega)\mathcal{G}^{\dag}_{1,3}] ,
\eea
where $\mathcal{G}$ is the phonon Green function across the molecule and the spectral function $\Lambda$ is defined as
\bea
\Lambda_{L,R}(\omega) & \equiv &  i[\Sigma^{ph}_{L,R}-\Sigma^{ph\;\dag}_{L,R}] 
= \omega^2_0 \sin Ka \cr
&=& \sqrt{\omega^2(2\omega^2_0-\omega^2)}.
\eea
Here the phonon self energy $\Sigma^{ph}_{L,R}$ due to the leads, in analogy with the electron self energy given in Eq.~(\ref{Sigma-el}), is given by  
\bea
\Sigma^{ph}_{L,R}=-\frac{\omega^2_0}{2}e^{iKa}.
\label{ph-self-energy}
\eea

If $\xi(E)=1$ for all $E$, then the equilibrium linear response contribution can be obtained exactly, showing that each massless phonon mode contributes a quantum of $\pi^2/3$ to the energy flux \cite{rego}. As a comparison, using the parameters in Figure \ref{fig-TofE-SingleChain}, we get the number current $I_N \approx 0.1$ and $I_e\approx 0.14$; adding $\pi^2/3$ to $I_e$ would reduce the efficiency from $\eta/\eta_c\approx 0.4$ to $\eta/\eta_c\approx 0.03$. However, it is also clear that the phonon transmission can be significantly reduced by designing the central `molecule' to have a large mass compared to the atoms in the lead. 
\begin{figure}[!tp]
\begin{center}
\includegraphics[angle=0, width=0.3\textheight]{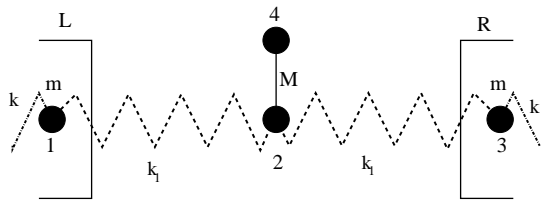}
\end{center}
\caption{The single chain phonon model where the `molecule' $2-4$ is assumed to have a mass $M$, connected by two springs, each of spring constant $k_1$, to the surface sites in the two leads each having mass $m$. Each surface mass $m$ is connected to its nearest neighbor in the lead by a spring with a spring constant $k$. The bulk of the leads are also made of masses $m$ connected by springs of spring constant $k$.}
\label{fig-toy-phonon-model}
\end{figure}
Indeed, for an order of magnitude estimate,  let us reconsider the single-chain model, with the $V$-sites replaced by `balls' of mass $m$ and the $t$-bonds replaced by `springs' of spring constant $k$ in the left and right semi-infinite leads as shown in Figure \ref{fig-toy-phonon-model}. The `molecule' of sites $2-4$ is replaced by a ball of mass $M$ and the bonds $t_1$ are replaced by springs of spring constant $k_1$. The transmission function will then be determined by the frequencies $\Omega^2_0\equiv 2k_1/M$ and $\Omega^2_1\equiv 2k_1/m$. It is easy to estimate that e.g. for $k_1=k$ and $M/m=10$ such that $\hbar\Omega_1=20 k_BT_L$ and  $\hbar\Omega_0=2 k_BT_L$ where $k_BT_L=0.5$, the phonon contribution is $I_{ph} \approx 0.0003$, which is negligible compared to the electronic contribution $I_e\approx 0.14$. Note that just as in the electronic case, we do not expect the efficiency to decrease if this one-chain contribution scales with increasing number of chains.

\section{Summary and conclusion}

As noted by a number of authors, nanostructured thermoelectric devices constructed from molecular juntions show potential because the transmission coefficient near the Fermi energy can be rapidly varying.  This leads to
a large thermoelectric figure of merit, $ZT$.  In this paper we show that there are other necessary conditions for these devices to be useful besides just a high thermoelectric figure of merit.  Specifically, we have addresed three questions.

First, because the length scales are so small in nanostructures, any non-infinitesimal temperture or electrical potential difference leads to very large gradients.  Nanostructures and molecular junctions in particular are inherently in the nonlinear response regime. Starting in Sec. II we show that just having a rapidly varying transmission
coefficient near the Fermi energy is not sufficient to get a large efficiency or power output in the nonlinear response regime. Rather in the nonlinear response regime, the crucial factor is where the transmission coefficient is weighted relative to the difference in the Fermi distribution functions of the leads.  The optimum transmission coefficient depends on the temperature and the chemical potential difference.

Second, we are ultimately interested in the power output from a thermoelectric
device.  While for ordinary thermoelectric devices one would assume that 
the large power output occurs with high efficiency, we show that for some
tunneling models the efficiency approaches the thermodynamic maximum
as the power output goes to zero.  This raises the question of whether
it is possible to have both high power output and high efficiency in 
tunneling through molecules.  In Sec. III we consider a model system
that has been used by a number of authors including some who fit it
to microscopic calculations.  Within this model, called the $t$-stub,
we find that it is possible to have both high power output and large
efficiency.  Random sampling of the parameters in the model show that
this occurs for a range of parameters -- not just for some very specific ones.

Finally, because nanostructures are so small, the currents and energy
transfers are also quite small.  For any macroscopic device one would
like to scale up the response of individual nanostructure devices by having
many act in parallel.  However, when molecules are placed close enough
together, there will be tunneling between them and the single molecule
calculations are no longer valid.  Thus, in Sec. IV we calculate the
transmission coefficient, efficiency, and power for many coupled
$t$-stub devices in parallel.  We find that the coupling does not necessarily
destroy the effect, and it is still possible to obtain both high efficiency and 
high power.  We estimate the power produced by many t-stub devices in
parallel and find that they could in principle be commerically viable.

Thus, thermoelectric devices constructed by tunneling through a molecule
are still promising upon closer inspection.  At least in one model
which has been mapped onto realistic systems, it is possible to obtain
high power output, high efficiency, and also to scale up the response
by placing many coupled nanostructure devices in parallel.

Nonetheless, we have made several common approximations that need to
be addressed in future more realistic calculations.  We have assumed that
the hot electrons dissipate their energy in the leads
and that energy is carried away rapidly.  This is a common assumption
in molecular tunnel junctions.  This energy dissipation should ultimately
be modeled microscopically with phonons or inelastic electron-electron
scattering on the molecule or in the leads. We have included in Sec V phonons 
to carry heat between the two leads  but not interacting with the electrons. 
Including scattering with the electrons would address potential
heating issues and also any loss of coherence effects caused by inelastic
scattering.  It would also address the effects of the Coulomb interaction
beyond the average effects included in static electronic structure
calculations. Future work will include some of these inelastic 
scattering mechanisms microscopically.

\section*{Acknowledgments:} KAM is grateful to J-L Pichard for introducing him to the current issues in thermoelectricity during a sabbaatical stay at Saclay, France, supported in part by RTRA Triangle de la Physique (Project Meso-Therm).

\section{Appendices}

\subsection*{Appendix I: Linear response regime and Figure of Merit}

In order to make connection with the figure of merit in the linear response regime, 
let us start by expanding the function 
\bea
F(E) & \equiv & f_L(\mu_L,T_L; E)-f_R(\mu_R,T_R; E); \cr 
f_j(\mu_j, T_j; E) &\equiv & 1/(1+e^{(E-\mu_j)/k_BT_j})
\eea
for small chemical potential difference $\mu_L-\mu_R$ and small temperature difference $T_L-T_R \equiv \Delta T$:
\bea
F(E) &\approx &  \left(-\frac{\partial f(E-\mu_{eq})}{\partial E}\right)\cr
&\times & \left((\mu_L-\mu_R)+(E-\mu_{eq})\frac{\Delta T}{T}\right)
\eea
where $T$ is the average temperature. The number and energy currents across the junction then becomes
\bea
I_N &=& (\mu_L-\mu_R)L_0+\frac{\Delta T}{T}L_1\cr
I_E &=& (\mu_L-\mu_R)L_1+\frac{\Delta T}{T}L_2,
\eea
where we have defined 
\bea
L_n\equiv \int dE\;(E-\mu_L)^n \mathcal{T}(E) \frac{-\partial f}{\partial E}.
\label{eq-Ln}
\eea
Here $L_0$ and $L_2$ are positive, but $L_1$ can be positive or negative, satisfying the relation $(L_2/L_0) > (L_1/L_0)^2$.

For the optimization of the efficiency, let us first consider the case when $I_N=0$. The `open circuit' chemical potential difference is 
\bea
\mu_L-\mu_R = -\frac{\Delta T}{T}\frac{L_1}{L_0} \equiv \Delta \mu_0.
\eea
Using $\Delta \mu_0$, we can rewrite
\bea
I_N &=& L_0 [(\mu_L-\mu_R)-\Delta \mu_0]\cr
I_E &=& L_1 [(\mu_L-\mu_R)-\Delta \mu_0(1+\delta)]
\eea
where 
\be
\delta \equiv \frac{L_2L_0}{L_1^2}-1 > 0.
\ee
The efficiency then becomes
\be
\eta= \frac{L_0 [(\mu_R-\mu_L)[(\mu_L-\mu_R)-\Delta \mu_0]}{L_1 [(\mu_L-\mu_R)-\Delta \mu_0(1+\delta)]}.
\ee
If $L_1 >0$, then $\Delta \mu_0 <0$ and $\mu_L-\mu_R <0$, and if $L_1 <0$, then $\Delta \mu_0 >0$ and $\mu_L-\mu_R >0$. However, in both cases the ratio $x\equiv (\mu_L-\mu_R)/ \Delta \mu_0$ is positive and in particular $0 < x < 1$. In terms of $x$ the efficiency becomes
\bea
\eta= \frac{L_0 [(-x\Delta \mu_0)(x-1)}{L_1(x- 1-\delta)}=\frac{\Delta T}{T}\frac{x(1-x)}{1+\delta-x}.
\eea
The figure of merit, defined as $ZT \equiv S_e^2T\sigma/\kappa$, is then identified with 
\be
ZT = \frac{1}{\delta}.
\label{eq-ZT}
\ee
Thus in the linear response regime, maximizing $ZT$  corresponds to minimizing $\delta$ and hence maximizing the efficiency. 

As an estimate, since the maximum efficiency $\eta_{max}$ occurs for $x\approx 1/2$, we have 
\be
\eta_{max}\approx \frac{\Delta T}{T} \frac{1}{4}\frac{1}{\delta+1/2}.
\ee
For $ZT=3=1/\delta $, we have
\be
\eta_{max}\approx 0.3 \frac{\Delta T}{T}. 
\ee

\subsection*{Appendix II: Proof of Eq. (\ref{GpkqkX})}

Here we give a proof of the equivalence of the two equations (\ref{GpkqkXY}) and (\ref{GpkqkX}).
Using the definitions of $Y_{m,k}$, we rewrite Eq. (\ref{GpkqkXY}) as
\be
G^n_{p_kq_k} =  g_0X_k\left[1+ X_k\sum_{j=1}^{n-k}b^j\prod_{i=k+1}^{k+j}X_{i}\prod_{m=k+1}^{k+j-1} X_{m}\right]
\ee
Now we use Eq. (\ref{solnXn}) to write
\bea
\prod_{i=k+1}^{k+j}X_{i} &=& 2^j \frac{(1+\alpha)^{k+1}-(1-\alpha)^{k+1}}{(1+\alpha)^{k+j+1}-(1-\alpha)^{k+j+1}} \cr
&=& \frac{2^j}{(1+\alpha)^j}\frac{1-z^{k+1}}{1-z^{k+j+1}}.
\eea
Then the Green function becomes
\bea
G^n_{p_kq_k} &=&  g_0X_k\left[1+ \sum_{j=1}^{n-k}\frac{(4b)^j}{(1+\alpha)^{2j}} \frac{1-z^{k+1}}{1-z^{k+j+1}} \frac{1-z^{k}}{1-z^{k+j}}\right]\cr
&=& g_0X_k\left[1+\sum_{j=1}^{n-k} \frac{z^j(1-z^k)(1-z^{k+1})}{(1-z^{k+j})(1-z^{k+j+1})}\right]
\eea
where we have used
\be
\frac{4b}{(1+\alpha)^2}=\frac{1-\alpha^2}{(1+\alpha)^2}=\frac{1-\alpha}{1+\alpha}=z.
\ee
We rewrite
\bea
&\;& \frac{z^j}{(1-z^{k+j})(1-z^{k+j+1})} \cr 
&=&  \frac{1}{z^k(1-z)}\left[\frac{1}{1-z^{k+j}}- \frac{1}{1-z^{k+j+1}}\right],
\eea
giving
\bea
G^n_{p_kq_k} &=&  g_0X_k\left[1+ Z_k  \sum_{j=1}^{n-k}\left(\frac{1}{1-z^{k+j}}- \frac{1}{1-z^{k+j+1}}\right)\right]\cr
&=& g_0X_kZ_k \sum_{j=0}^{n-k}\left[\frac{1}{1-z^{k+j}}- \frac{1}{1-z^{k+j+1}}\right]
\eea
where we defined 
\be
Z_k\equiv \frac{(1-z^k)(1-z^{k+1})}{z^k(1-z)}
\ee
and we have extended the sum from $j=0$ to include the first term equal to 1. 

Note that in the difference of the two sums, only the $j=0$ contribution of the first term and $j=n-k$ of the second term survive, the rest canceling each other. This gives 
\bea
&\; & \sum_{j=0}^{n-k} \left[\frac{1}{1-z^{k+j}}- \frac{1}{1-z^{k+j+1}}\right] = \frac{1}{1-z^{k}}- \frac{1}{1-z^{n+1}}\cr
&=& \frac{z^k(1-z^{n-k+1})}{(1-z^k)(1-z^{n+1})}.
\eea
Using 
\be
X_k=(1+z)\frac{1-z^k}{1-z^{k+1}}
\ee
we finally get
\bea
G^n_{p_kq_k} &=&  g_0(1+z)\frac{1-z^k}{1-z^{k+1}} \frac{(1-z^k)(1-z^{k+1})}{z^k(1-z)}\cr &\times &\frac{z^k(1-z^{n-k+1})}{(1-z^k)(1-z^{n+1})}\cr
&=& g_0\frac{1+z}{1-z}\frac{(1-z^k)(1-z^{n-k+1})}{1-z^{n+1}}.
\eea

This is identical to Eq. (\ref{GpkqkX})

\subsection*{Appendix III: Continuum model}

The on-site (retarded) Green function $G_{22}$ for a single wire ($t_0=0$) connected to the leads is given by $G^R_W\equiv G_{22}=E_0a_0/D_1$ (compare with Eq.~(\ref{G13}) for $G_{13}$). When wires are connected by a hopping matrix element $t_0$ at the center to form a half space, the Green function at the central site is
\bea
t_0G^R_H = \frac{(G^R_W)^{-1}}{2t_0}\pm i\sqrt{1-\left(\frac{(G^R_W)^{-1}}{2t_0}\right)^2}.  
\eea
The physical situation is where the imaginary part of the inverse of the retarded Green function is positive. The full Green function of the entire space of wires is 
\bea
(G^R_F)^{-1} = \pm i 2t_0 \sqrt{1-\left(\frac{(G^R_W)^{-1}}{2t_0}\right)^2}.
\eea
In terms of the parameters $g_0$ and $\alpha$ defined in Eqs.~(\ref{G13}) and (\ref{solnXn}), this can be rewritten as 
\bea 
(G^R_F)=\pm \frac{G^R_W}{\alpha}=\frac{t^2}{t^2_1}e^{-2ika}\frac{g_0}{\alpha}.
\eea
The transmission function in terms of this central site Green function is given by
\bea
\mathcal{T'}(E)=\frac{t^2_1}{t}\sin (ka) i(G^R_F-G^A_F)
\eea
where $G^A$ is the advanced Green function. Using the expression for $G^R_F$ above, and the fact that $G^A$ is the complex conjugate of $G^R$, we finally obtain Eq.~(\ref{Tinfinity}).

\end{document}